\documentclass[aps,prl,twocolumn,showpacs]{revtex4}

\newcommand{\Jeff}{J_{\mbox{\footnotesize eff}}}

\usepackage{graphicx}

\begin{document}

\title{Directed transport in driven optical lattices
by gauge generation}

\author{C.E.~Creffield and F.~Sols}
\affiliation{Departamento de F\'isica de Materiales, Universidad
Complutense de Madrid, E-28040, Madrid, Spain}

\date{\today}


\begin{abstract}
We examine the dynamics of ultracold atoms held in optical lattice
potentials. By controlling the switching of a
periodic driving potential we show how
a phase-induced renormalization of the intersite tunneling
can be used to produce directed motion and control wavepacket
spreading. We further show how  
this generation of a synthetic gauge potential
can be used to
split and recombine wavepackets, providing an attractive route
to implementing quantum computing tasks.
\end{abstract}

\maketitle

{\em Introduction -- }
In recent years enormous experimental progress has been made in creating
and trapping ultracold atom gases \cite{morsch_review}. 
When placed in an optical
lattice potential these gases provide extremely clean
and controllable implementations of interacting lattice systems, since 
parameters such as the interparticle interaction, the
lattice depth and spacing are all readily tunable. Dissipation
and decoherence effects are typically extremely weak, allowing the 
quantum coherent behaviour of these systems to be directly observed. 

In contrast to electronic systems, however, trapped atoms are uncharged,
and so electric or magnetic fields cannot easily be used to produce or
regulate transport. 
Due to their excellent coherence properties, 
one means of controlling the dynamics of the atoms is via
quantum interference effects. A notable example is termed 
``coherent destruction of tunneling'' (CDT) in which 
a periodic driving of the lattice causes
the amplitude of the intersite hopping to be renormalized \cite{cdt}.
This renormalization has been seen directly in
the expansion of trapped Bose-Einstein condensates
\cite{pisa_flat, pisa_tilt, alberti}, and has been used very recently 
to produce the fascinating phenomenon
of ``super Bloch oscillations'' \cite{alberti,superbloch}, and
to induce the quantum phase
transition \cite{pisa_mott} between a 
superfluid and an insulator. 

In this Letter we show that as well as controlling the amplitude of 
the hopping, a periodic driving field can also be used to
produce a tunneling {\em phase}, equivalent to a $U(1)$ gauge potential.
This gauge potential arises
from the combined effect of the phase of the driving field, and the
careful control of the switching condition.
Although here we only consider one-dimensional lattices, the technique
can also be similarly applied to create hopping phases in higher-dimensional 
systems. In this case the phases can be interpreted as Aharonov-Bohm
phases picked up by a particle hopping from site to site, corresponding
to a synthetic magnetic field threading the lattice \cite{kolovsky}.
Other schemes have been devised to produce 
such gauge potentials in cold atom systems, 
including lattice rotations, 
state-dependent optical potentials \cite{flux},
or phase-imprinting \cite{imprint}. Our procedure, however,
has an appealing simplicity requiring only the periodic
vibration of the lattice potential, which is easily produced
in experiment. We show how the driving can be used to control 
both the spreading and position
of an initial wavepacket, and in particular, how a {\em directed} current
of {\em non-dispersing} wavepackets can be induced.
We shall also demonstrate 
how wavepackets can be split, guided, and
recombined in a controllable and robust manner,
accessible to current experiment.

{\em Model -- }
A gas of weakly interacting ultracold bosonic atoms
can be described well by the Gross-Pitaevskii
equation (GPE). When a sufficiently deep optical lattice
potential is applied, the wavefunction will localize
mainly in the potential minima defining the
lattice sites, making it convenient 
to use a discretised form of the GPE
\begin{equation}
i \frac{\partial \psi_j}{\partial t} = - \left( J \psi_{j+1} +
J^\dagger \psi_{j-1} \right) + g \left| \psi_j \right|^2 \psi_j  + 
j V(t) \psi_j .
\label{gpe}
\end{equation}
Here $\psi_j$ denotes the system's wavefunction on lattice site $j$, 
and $J$ describes the
the tunneling amplitude between nearest neighbor sites.
Interactions between the bosons are given by a
mean-field interaction, set by the nonlinearity parameter $g$. 
The time-dependent driving potential is assumed to
rise linearly across the lattice \cite{pisa_flat, pisa_tilt, superbloch},
and has a time dependence given by
$V(t) = \Delta +  K \sin(\omega t + \phi)$, where
$\Delta$ is a static tilt of the lattice potential,
and $\omega$ and $K$ are the frequency and amplitude respectively
of the oscillating component.

As an initial state we take a Gaussian wavepacket, 
$\psi_j = N \exp\left[- j^2/2 \sigma_0^2 + i \theta_j\right]$,
where $\sigma_0$ is the initial width of the wavepacket measured in units
of the lattice spacing, and $N$ normalizes the wavefunction to unity.
This choice of initial state mimics the experimental
situation \cite{pisa_flat, pisa_tilt, superbloch}, in which the
condensate is prepared in a harmonic trap,
and so typically has a Gaussian profile when transferred to
the optical lattice. Note also that we explicitly include a
site-dependent phase term $\theta_j$ in the
wavefunction.

{\em Analysis -- }
We first consider the non-interacting case ($g=0$). The Hamiltonian
describing the system (\ref{gpe}) is then $T$-periodic in time, 
where $T = 2 \pi/\omega$, and the
natural framework to describe its time evolution is Floquet theory.
This reveals that in the high-frequency limit ($\omega > J$)
the time-dependent driven system can be 
described by an effective {\em static} Hamiltonian, whose parameters can be
systematically evaluated by using perturbation theory \cite{holthaus}
on the Floquet states. 
While these states are explicitly time dependent, being $T$-periodic
functions, in the high-frequency limit their time variation
is rather weak. This is the origin, for example, of the well-known
negligible time-dependence of CDT \cite{cdt}, as compared to the 
large oscillations
observed at low driving frequencies when dynamical localization instead
occurs \cite{kenkre}.

To first-order the tunneling amplitudes are
modified as $J \rightarrow
\Jeff = J \langle \exp \left[- i \int_0^t V(t') dt' \right] \rangle$,
where $\langle \cdots \rangle$ indicates a time-average over the driving
period $T$. We restrict ourselves to considering the case of
{\em resonant driving}, when $\Delta = n \omega$, which 
yields the result
\begin{equation}
\Jeff / J = e^{-i K / \omega \cos \phi}
e^{i n \left( \phi + \pi/2 \right) } 
{\cal J}_n\left(K / \omega \right) ,
\label{effective}
\end{equation}
where ${\cal J}_n$ is the $n$th Bessel function of the first kind.
For the case of $n = 0$, a
similar expression was obtained in Ref.\onlinecite{cec_sols_2008}
for a driven Bose-Hubbard model.
There a ramped driving potential was used
to adiabatically transform the ground state of the Hamiltonian
to a stroboscopically current-carrying Floquet state. Here, however,
we use the tunneling phase in a very different way, to
control the {\em non-equilibrium}
dynamics of an expanding atomic wavepacket.
This, in conjunction with the weak time-dependence
of the Floquet states in the high-frequency regime, means that
the results we present are {\em not} stroboscopic, and
have a negligible dependence on the moment within each driving
period at which the system is measured.

From Eq.(\ref{effective}) we can immediately note the importance of the phase
of the driving, $\phi$. For a cosinusoidal driving ($\phi = \pm \pi/2$),
the most frequently considered case in the literature, this result simplifies
to yield $\Jeff / J= (-1)^n {\cal J}_n\left( K / \omega \right)$ -- 
the well-known Bessel function renormalization of tunneling found in CDT.
For {\em sinusoidal} driving ($\phi=0$), however, the tunneling
additionally acquires a phase
$\Jeff / J= \exp\left[-i \left(K/\omega - n \pi/2 \right) \right] 
{\cal J}_n\left( K / \omega \right)$.

It is natural to ask whether this tunneling-phase has physical 
implications, since it would appear that
$\phi$ can simply be eliminated by
a shift of the time coordinate. It is important to note, 
however, that we consider the driving potential $V(t)$ to be switched on
at a specific moment $t=0$, in common with 
experimental implementations \cite{alberti,superbloch}.
This gives the time origin, and thus the driving-phase, 
an unambiguous definition, and consequently $\phi$ can
indeed be of experimental relevance, as 
noted in Ref.\onlinecite{tania}.
This differs from many theoretical analyses \cite{rivals}, in 
which the {\em steady-state} properties of a driven system are
considered, and the driving is implicitly assumed to have been
turned on at $t \rightarrow -\infty$.
In such cases the phase of the driving is indeed unimportant.

The expansion of an initially Gaussian condensate in
a periodically-driven lattice was analyzed in Ref.\onlinecite{pra}
for real values of $\Jeff$.
Extending this analysis to complex $\Jeff$ gives the result 
\begin{equation}
\sigma(t) =  \sigma_0 
\sqrt{ 1 + \left( \Re[\Jeff] t / \sigma_0^2 \right)^2} ,
\label{expand}
\end{equation}
We thus see that
the {\em spreading} of the wavepacket is governed by
the {\em real} component of $\Jeff$. For $n = 0$, for example,
$\Re[\Jeff] = {\cal J}_0 \left(K / \omega \right) 
\cos \left( K / \omega \ \cos \phi \right)$, and so as well
as freezing at the ``standard'' CDT condition (when the
Bessel function vanishes), expansion is also suppressed at an additional
set of values where $\cos \left( K / \omega \ \cos \phi \right)= 0$.

As well as the expansion of the condensate, another useful
experimental measurement is its center of mass motion.
In the absence of driving our system has 
the standard spectrum of a non-interacting lattice model,
$E_k = -2 J\cos k$. When the system is driven, we can
replace the energies $E_k$ with quasienergies, obtained as solutions
of the Floquet equation, to obtain the new dispersion relation
$\varepsilon_k = -2 |\Jeff| \cos(k - k_0)$, where 
$\Jeff = |\Jeff| \exp[i k_0]$. The effect of the tunneling phase
is thus not to alter the quasienergy spectrum of the system,
but to displace the wavepacket to another point in the first 
Brillouin zone. In analogy with the familiar semiclassical expression
we can now define a mean group
velocity, $\overline{v}_g = d \varepsilon_k / d k$,
where the average is taken over one period of the driving,
to obtain the final result
\begin{equation}
\overline{v}_g = - 2 \Im[\Jeff] .
\label{velocity}
\end{equation}
We thus arrive at the rather elegant result that the
two quantities most accessible to experiment -- the wavepacket
expansion and its center of mass motion -- are directly
related to the real and imaginary parts respectively of $\Jeff$.

\begin{center}
\begin{figure}
\includegraphics[width=0.4\textwidth,clip=true]{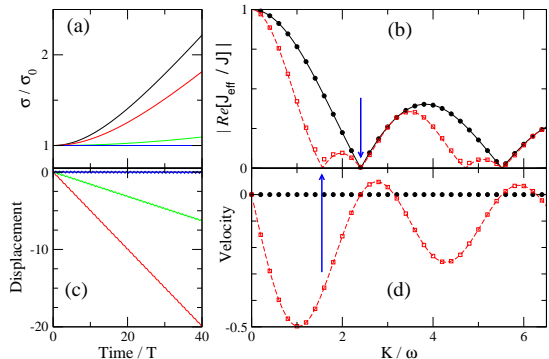}
\caption{Response of a Gaussian wavepacket ($\sigma_0 = 4$) to a 
periodic driving potential, $\omega = 16 J$.
(a) Wavepacket expansion under cosinusoidal driving; from top
to bottom $K/\omega = 0, 1, 2,2.4$. For $K/\omega=2.4$ CDT 
occurs, and the expansion is suppressed.
(b) Real component of $\Jeff$, extracted from the expansion curves for
cosinusoidal (black circles) and sinusoidal (red squares) driving.
The curves show the theoretical prediction 
obtained from Eq.\ref{expand}.
(c) Displacement of a wavepacket under sinusoidal driving 
in units of the lattice spacing.
For $K/\omega=0$ and $2.4$ no displacement occurs; otherwise
it increases linearly with time.
(d) As in (b), for the wavepacket velocity, given 
in units of $d_L / T$ where $d_L$ is the lattice spacing.
Vertical blue arrows mark the driving parameters $K/\omega=\pi/2$
(dispersionless directed transport) and $K/\omega=2.404$
(complete suppression of dynamics).}
\label{driving}
\end{figure}
\end{center}

{\em Directed transport -- }
To verify these results we numerically
simulate the model (\ref{gpe}) for a $200$ site lattice 
with no static tilt ($\Delta = 0$),
and take the onsite phases $\theta_j$ to be constant.
In Fig.\ref{driving}a
we show the condensate's expansion for a cosinusoidal driving
for several values of $K/\omega$. These curves consist of an initial
quadratic dependence on $t$ followed by a linear ballistic expansion
at long times \cite{pra}, and clearly show how
varying $K/\omega$ controls the condensate spreading. 
Eq.\ref{expand} can be used to extract the value of 
$| \Re [ \Jeff ] |$ from these expansion curves, which we
plot in Fig.\ref{driving}b. The expected Bessel function dependence
of $\Jeff$ is clearly seen,
with the condensate expansion being frozen at the
zeros of ${\cal J}_0$ (for $K/\omega = 2.40, \ 5.52 \dots$).
However, the corresponding expansion for sinusoidal driving shows
an additional set of zeros at $K/\omega=\pi/2, 3 \pi/2 \dots$,
in exact agreement with 
Eq.\ref{effective} for $n=0$.
At these values of driving
the suppression of the expansion arises from a very different cause;
the tunneling phase displaces the wavepacket 
to $k_0 = \pi/2$ in the first Brillouin
zone where the quasienergy bands have an inflexion,
causing the effective mass to diverge
and so quenching the spreading of the wavepacket.

In Fig.\ref{driving}c we show the motion of the center of mass
of the condensate under sinusoidal driving.
The wavepacket, initially at rest,
begins to move at a constant rate,
depending on $K/\omega$. In Fig.\ref{driving}d we plot the
velocity corresponding to this displacement, and find that it agrees
excellently with the predicted mean group velocity ${\overline v}_g$
(Eq.\ref{velocity}). Under cosinusoidal driving, however, the
velocity of the wavepacket is zero, also as predicted.

We note that to obtain Eqs.\ref{expand},\ref{velocity} we have assumed that
the driving field can be turned on instantaneously. In experiment,
of course, this idealized behaviour is not possible.
To check if our results are robust to this effect, we have included 
a ramp function in $V(t)$ to describe the effect of turning the field 
on from zero during a short, but finite, time interval. We find that as long
as the ramp-time is sufficiently short, $\lesssim 0.02 T$, very
similar results are obtained. Experiments typically use driving frequencies
of the order of kHz, which would thus demand ramp-times of $\sim 10 \mu$s,
which are achievable. 

{\em Directed motion -- }
We can thus see that $\phi$ can be used to 
cause an initially stationary wavepacket to move
in a given direction with a precisely defined velocity,
without requiring the spatial symmetry of the
lattice to be broken.
Two values of $K/\omega$ are of particular interest,
and are marked in Fig.\ref{driving}b,d.
For $K/\omega=2.404$ (the first zero of ${\cal J}_0$)
the expansion of the initial wavepacket is suppressed, and its
induced velocity is zero for all values of the driving
phase. This amounts to a complete suppression of the dynamics
of the condensate. However, at $K/\omega=\pi/2$ a wavepacket that
is sinusoidally driven will not expand, but will have a non-zero
velocity -- a {\em directed current} of {\em non-dispersive} wavepackets.
 
\begin{center}
\begin{figure}
\includegraphics[width=0.4\textwidth,clip=true]{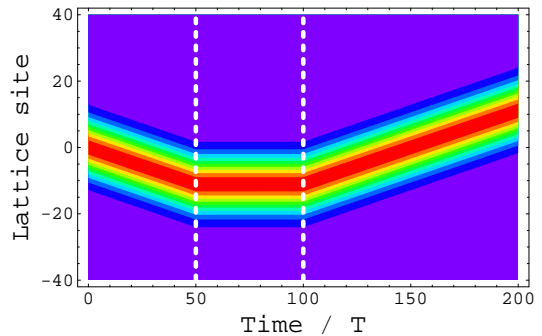}
\caption{Density plot of the motion of a wavepacket under sinusoidal driving. 
Initially $K/\omega=\pi/2$ to induce dispersionless transport; $K/\omega$
is then set to $2.404$ to freeze the motion, and finally to $-\pi/2$
to reverse it. The vertical dotted lines indicate the times
at which $K/\omega$ is changed.}
\label{control}
\end{figure}
\end{center}

In Fig.\ref{control} we show the motion of such a wavepacket. Initially
we set $K/\omega=\pi/2$ to induce motion. The driving
is then tuned to $K/\omega=2.404$ to bring the wavepacket to a halt,
and then to $K/\omega=-\pi/2$  to move the wavepacket in the opposite
direction. It is clear that the spreading of the wavepacket
is negligible, and that this technique indeed gives excellent
control over the system.
It is interesting to note that a similar form of control
was reported in Ref.\onlinecite{tino} for an {\em amplitude}
modulated lattice, instead of the phase-modulated lattice we consider.
An important difference between the two cases, however, is that
phase modulation does not require the presence of a static lattice
tilt, since the effects also occur for $n=0$,
whereas amplitude modulation is limited to the case of resonant driving ($n > 0$).
In addition, amplitude modulation does not produce CDT, the intersite
tunneling depending linearly on the driving amplitude instead of the Bessel function
dependence given in Eq.\ref{effective}. 

{\em Wavepacket splitting -- }
We now consider the effect of the onsite phases $\theta_j$.
It is well-known \cite{imprint} that imprinting a wavepacket with a
uniform phase gradient $\theta_{j+1} - \theta_j = \theta$
has the effect of inducing motion of the center of mass,
similar to the motion we have observed by manipulating
the driving-phase $\phi$. By simulating the system with different
values of $\theta$, we have confirmed that the two phases combine, so that
the net motion of the wavepacket actually depends on the phase difference 
$\phi - \theta$. The driving field can thus be used to separate
components of a wavepacket which possess different phase gradients.
Let us consider the case of a superposition of a wavepacket
with uniform phase ($\theta = 0$) and one with $\pi$-phase 
($\theta = \pi$). If the components have equal weight, the superposition
will have the form $\psi_j=N \exp[-j^2/2 \sigma_0^2 ]$ for $j$ odd,
and $\psi_j = 0$ for $j$ even (with no loss of generality we can
interchange the roles of the odd and even sites).
Such a state can be prepared, for example, by patterned loading of a
single uniform-phase condensate.

\begin{center}
\begin{figure}
\includegraphics[width=0.4\textwidth,clip=true]{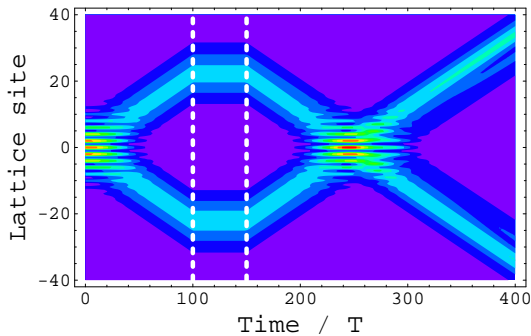}
\caption{Density plot of a two-component wavepacket under sinusoidal driving.
Initially $K/\omega=\pi/2$ and the wavepacket splits apart. Its motion is then
halted and reversed as before to bring about a collision. The interaction, $g=0.5 J$,
causes the final state to be asymmetric.}
\label{split}
\end{figure}
\end{center}

Under sinusoidal driving, for $K/\omega = \pi/2$, the component with
uniform phase will move, without distortion, at a negative velocity,
while the $\pi$-phase component will move identically but with a positive
velocity. The initial wavepacket will thus split apart, as shown
in Fig.\ref{split}. Tuning $K/\omega$ to $2.404$ will bring each
component to a stop, and then setting $K/\omega=-\pi/2$ will bring the
wavepackets together. For zero interaction ($g=0$) the wavepackets
will simply pass through each other. For small values of
$g$ the splitting process occurs as before, but during the collision
the interaction causes the wavepackets to distort and produces
a slightly asymmetric final state as shown.

\begin{center}
\begin{figure}[h]
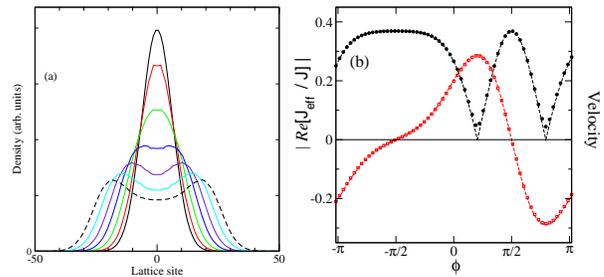

\includegraphics[height=0.2\textwidth,clip=true]{fig4a}
\includegraphics[height=0.2\textwidth,clip=true]{fig4b}
\caption{(a) Expansion of a phase incoherent wavepacket, obtained by averaging
over $200$ random realizations of the onsite phases. 
Driving parameters: $n=1$, $K/\omega=0.8$. The density
profile is shown in timesteps of $24 T$, showing the formation of a
double peak structure as the wavepacket spreads. 
(b) Symbols show results obtained from simulations for
the real component of the effective tunneling,
$\Re[\Jeff]$ (black circles), which controls
the rate of the wavepacket spreading, and 
velocity (red squares), given in units of $d_L / T$;
dashed lines show the analytical results. The minimum values of
$\Re[\Jeff]$ align with the maximum velocity, and vice-versa.}
\label{random}
\end{figure}
\end{center}

{\em Incoherent expansion -- }
An intriguing result seen in Ref.\onlinecite{superbloch} is that unlike
previous driven lattice experiments \cite{pisa_flat,pisa_tilt},
the wavepacket {\em deformed} under resonant driving, developing pronounced
edges during its expansion. As a possible explanation of this effect
we now look at the expansion of a phase incoherent wavepacket, by 
averaging over many realizations of random
onsite phases $\theta_j$. The result in Fig.\ref{random}a is
strikingly similar to the experimental observation.

The phase effects we have discussed give a simple explanation
of this behavior. A phase incoherent wavepacket can be expressed
as a mixture of many wavepackets, each with a random, but constant,
phase gradient $\theta$. Under the periodic driving each component will
both develop a certain velocity and will spread, according to 
Eqs.\ref{expand},\ref{velocity}. 
As the components which spread least
have the highest velocity, while those that move more slowly  
spread more rapidly (see Fig.\ref{random}b), 
the initial state will segregate
with the rapidly moving components at the edges of the wavepacket
remaining taller and narrower than the slower-moving components near the center.
The edges of the wavepacket will move at the maximum speed,
which for $n=1$ is given by
$v_{\mbox{\footnotesize max}}= |2 {\cal J}_1(K/\omega) |$.
For the driving parameters used, our model predicts 
$v_{\mbox{\footnotesize max}}= 847 \ d_L/$s, where
$d_L$ is the lattice spacing, which compares well with the
experimentally measured value of $869 \ d_L/$s. 
We thus suggest that the unusual expansion seen in 
Ref.\onlinecite{superbloch}
is a consequence of the phase incoherence of the initial state, possibly
arising from phase randomization produced by Wannier-Stark
localization during the preparation of the system.

{\em Conclusions -- }
We have shown how the phase of a driving potential
can be used to control the dynamics of an atomic wavepacket,
both by regulating its rate of expansion, and by inducing a
steady drift of its center of mass. Combining these effects 
allows the directed transport of non-dispersive wavepackets.
Periodic driving also acts as a ``prism'' for the separation
of different phase contributions within a wavepacket.
This allows wavepackets to be divided and recombined,
and also provides an appealing explanation for the unusual 
condensate expansion observed in Ref.\onlinecite{superbloch}.
While these results have been obtained within mean field theory,
probing the behavior 
of systems in the strongly-correlated regime remains an interesting 
subject for future research, holding out the enticing prospect of using
these effects to generate and distribute entanglement
in coherent lattice systems.
We also note that since these directed currents require coherence
across many lattice sites and driving cycles, their
eventual decay may provide information on decoherence mechanisms.

The authors thank Oliver Morsch for many
stimulating discussions, and acknowledge support
from the Spanish MICINN through
Grant Nos. FIS2007-65723, FIS2010-21372,
Acci\'on Integrada HI2008-0163,
and the Ram\'on y Cajal program (CEC).

\end{document}